\input psfig.sty
\documentstyle{cupconf}

\ifoldfss
\else
  \ifnfssone
    \newmathalphabet{\mathit}
      \addtoversion{normal}{\mathit}{cmr}{m}{it}
      \addtoversion{bold}{\mathit}{cmr}{bx}{it}
    \newmathalphabet{\mathcal}
      \addtoversion{normal}{\mathcal}{cmsy}{m}{n}
    \else
    \ifnfsstwo
    \fi
  \fi
\fi

\def\hexnumber#1{\ifcase#1 0\or1\or2\or3\or4\or5\or6\or7\or8\or9\or
 A\or B\or C\or D\or E\or F\fi }

\makeatletter
\ifx\CUP@mtlplain@loaded\undefined
\else
\fi
\makeatother
 \makeatletter
 \ifx\CUP@mtlplain@loaded\undefined
   \font\tenbmi=cmmib10 at 10pt
   \font\sevenbmi=cmmib10 at 7pt
   \font\fivebmi=cmmib10 at 5pt

   \newfam\bmifam
   \textfont\bmifam=\tenbmi
   \scriptfont\bmifam=\sevenbmi
   \scriptscriptfont\bmifam=\fivebmi
   
 \fi
 \makeatother
\ifnfsstwo

\fi
\ifnfssone

\fi
\ifoldfss

\fi

\mathchardef\varLambda="0103

\makeatletter
\ifx\CUP@mtlplain@loaded\undefined
\else
\fi
\makeatother
\makeatletter
\ifx\CUP@mtlplain@loaded\undefined
  \font\tenbms=cmbsy10
  \font\sevenbms=cmbsy10 at 7pt
  \font\fivebms=cmbsy10 at 5pt
  \newfam\bmsfam
  \textfont\bmsfam=\tenbms
  \scriptfont\bmsfam=\sevenbms
  \scriptscriptfont\bmsfam=\fivebms

  \edef\bsy@{\hexnumber\bmsfam}
  \mathchardef\bnabla="0\bsy@72
\fi
\makeatother
\title[Polarization Limits in CSOs]{Polarization Limits in Compact Symmetric Objects}

\author[Peck \& Taylor ]{%
A.\ns B.\ns P\ls E\ls C\ls K$^1$\ns \and \ns G.\ns B.\ns T\ls A\ls Y\ls L\ls O\ls R$^2$}

\affiliation{$^1$MPIfR, Auf dem H\"ugel 69, D-53121 Bonn, Germany\\[\affilskip]
$^2$NRAO, P.O. Box O, Socorro, NM 87801, USA} 

\begin{document}
\ifnfssone
\else
  \ifnfsstwo
  \else
    \ifoldfss
      \let\mathcal\cal
      \let\mathrm\rm
      \let\mathsf\sf
    \fi
  \fi
\fi

\maketitle

\begin{abstract}
We present results of multifrequency polarimetric VLBA observations of
34 compact radio sources.  The observations are part of a large survey
undertaken to identify CSOs Observed in the Northern Sky (COINS).
Based on VLBI continuum surveys of $\sim$2000 compact radio sources,
we have defined a sample of 52 CSOs and CSO candidates.  Positive
identification of CSOs is contingent upon acquiring multi-frequency
observations in order to correctly identify the core of the source,
which is expected to have a strongly inverted spectrum.  We also
expect CSOs to exhibit very little polarized flux.  Despite the fact
that synchrotron emission is intrinsically highly polarized, less than
0.5\% fractional polarization is seen in low resolution studies of
CSOs at frequencies up to 5 GHz.  One possible explanation for the low
observed linear polarization from CSOs is that their radiation is
depolarized as it passes through a magnetized plasma associated with a
circumnuclear torus.  This interpretation is consistent with the
unified scheme of AGN, and also with the recent detections of free-free
and H\kern0.1em{\sc i}\ absorption in CSOs.

Here we present limits on the polarized flux density at 8.4 GHz and
$\sim$1 mas resolution for 21 CSOs and candidates in the COINS sample.  

\end{abstract}

\firstsection 

\section{Introduction}
The 34 sources in this sample were initially selected from the VLBA
Calibrator Survey.  The total and polarized flux densities at 8 GHz of
the 21 sources which are considered to be CSOs (or remain CSO candidates
pending identification of the core) are shown in Table 1.  None
of these sources exhibit any significant polarization, with upper
limits ranging from 0.3\% to 1.6\%.

The fluxes of the sources which were rejected from the CSO list based
on morphology at 1.6, 5, 8 and/or 15 GHz and the spectral indices of
the components, as well as continuum images of all sources, can be
found in Peck \& Taylor (2000).  Five of the 13 rejected sources
exhibit measurable polarized emission at 8.4 GHz.

\renewcommand{\baselinestretch}{1.2}
\scriptsize
\begin{table}
\begin{center}
\begin{tabular}{ccccrr}
Source&RA & Dec &Opt. &$S_{8.4}$ &$P_{8.4}$ \\
Name&(J2000) &(J2000) & ID &(mJy) &(mJy) \\

J0000+4054 &00 00 53.0815&40 54 01.806&G &237.8&$<$1.2  \\
J0003+4807 &00 03 46.0413&48 07 04.134&...&80.6&$<$0.9 \\
J0132+5620& 01 32 20.4503&56 20 40.372& ...&291.1&$<$0.9 \\
J0204+0903& 02 04 34.7589&09 03 49.248&...&316.9&$<$1.2  \\
J0332+6753& 03 32 59.5241&67 53 03.860&...&...&... \\
J0427+4133& 04 27 46.0455&41 33 01.091&...&650.3&$<$0.9  \\
J0518+4730& 05 18 12.0899&47 30 55.536&...&344.8&$<$0.7  \\
J0620+2102& 06 20 19.5286&21 02 29.501&...&226.1&$<$0.9  \\
J0754+5324 &07 54 15.2177&53 24 56.450&...&73.8&$<$1.2  \\
J1111+1955& 11 11 20.0694&19 55 35.950&G &224.7&$<$0.9 \\
J1143+1834& 11 43 26.0706&18 34 38.375&... &233.9&$<$0.6 \\
J1311+1417& 13 11 07.8250&14 17 46.659&QSO&250.9&$<$0.9   \\
J1311+1658& 13 11 23.8204&16 58 44.213&...&244.3&$<$1.2  \\
J1414+4554& 14 14 14.8526&45 54 48.730&G & 76.1&$<$1.2 \\
J1415+1320&14 15 58.8188&13 20 23.714&QSO&1426.4&$<$4.5  \\
J1546+0026& 15 46 09.5312&00 26 24.615&G& 536.0&$<$0.9 \\
J1734+0926& 17 34 58.3773&09 26 58.274&G & 401.2&$<$1.2  \\
J1816+3457& 18 16 23.8987&34 57 45.729&G&272.7&$<$1.2  \\
J1826+1831& 18 26 17.7118&18 31 52.915&...&146.5&$<$1.5 \\
J2203+1007 &22 03 30.9534&10 07 42.584&...&172.7&$<$1.2 \\
J2245+0324 &22 45 28.2846&03 24 08.863&QSO& 390.3&$<$1.2\\

\end{tabular}
\caption{CSOs and CSO candidates in the COINS sample}
\end{center}
\end{table}
\normalsize

 \begin{figure}[!h]
 \centerline{\psfig{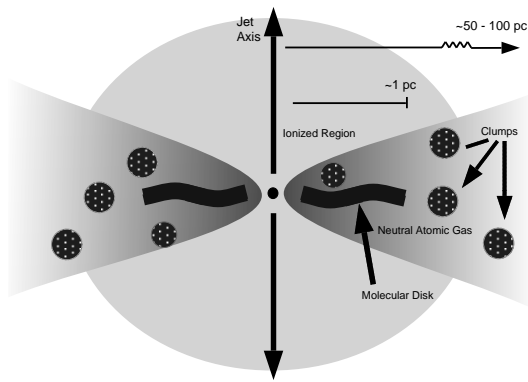}}
 \caption{A cartoon model of the central parsecs of a CSO.}
 \end{figure}

\section{Discussion}

Figure 1 depicts a cartoon of the possible environment in the central
parsecs of a CSO or other compact AGN.  This model is based on the
unified scheme, and on recent H\kern0.1em{\sc i}\ and free-free
absorption studies ({\it e.g.} Peck, Taylor \& Conway 1999).  Some
notable simplifications have been made for this figure.  For example,
it is not necessary for the toroidal structure to be perpendicular to
the jet axis, the ``clumps'' of denser gas are unlikely to be uniform
in size, and it is believed that the degree of warp in the disk can
vary greatly.  One possible explanation for the low observed levels of linear
polarization from CSOs is that our line of sight to the jet components
in these sources passes through a magnetized plasma associated with
this torus.  In order to depolarize the radio emission within our 8
MHz IF at 8.4 GHz the Faraday rotation measures could be larger than
5$\times$10$^5$ radians m$^{-2}$, or alternatively the magnetic fields
in the torus could be tangled on scales smaller than the telescope
beam of $\sim$1 mas to produce gradients of 1000 radians m$^{-2}$
mas$^{-1}$ or more.
\firstsection

\end{document}